\documentclass[aps,prb,twocolumn,superscriptaddress,showpacs]{revtex4-1}
\usepackage{graphicx}
\usepackage{calc}
\usepackage{bm}

\begin{document}

\title{A quantum Hall Mach-Zehnder interferometer far beyond the equilibrium}

\author{E.V.~Deviatov}
\email[Corresponding author. E-mail:~]{dev@issp.ac.ru}
\affiliation{Institute of Solid State Physics RAS, Chernogolovka, Moscow District, 142432, Russia}

\author{A.~Ganczarczyk}

\author{A.~Lorke}
\affiliation{Laboratorium f\"ur Festk\"orperphysik, Universit\"at Duisburg-Essen, Lotharstr. 1, D-47048, Duisburg, Germany}

\author{G.~Biasiol}
\affiliation{IOM CNR, Laboratorio TASC, 34149 Trieste, Italy}

\author{L.~Sorba}
\affiliation{NEST, Istituto Nanoscienze-CNR and Scuola Normale Superiore, 56127 Pisa, Italy}

\date{\today}

\begin{abstract}
We experimentally realize quantum Hall Mach-Zehnder interferometer which operates far beyond the equilibrium. The operation of the interferometer is based on allowed intra-edge elastic transitions within the same Landau sublevel in the regime of high imbalances between the co-propagating edge states. Since the every edge state is definitely connected with the certain Landau sublevel, the formation of the interference loop can be understood as a splitting and a further reconnection of a single edge state. We observe an Aharonov-Bohm type interference pattern even for low-size interferometers. This novel interference scheme demonstrates high visibility even at millivolt imbalances and survives in a wide temperature range. 
\end{abstract}


\maketitle

\section{Introduction}

Recent investigations of quantum Hall (QH) interferometers~\cite{heiblum,heiblum+06,heiblum+07,litvin,litvin1,glattli,roulleau,roulleau1,wiel,ofek,Camino+05,Camino+07a,Camino+07b,Ping+09,Zhang+09,McClure+09} give rise to several fundamental puzzles even in the integer QH regime. The major ones are the influence of the electron-electron interaction on the interference pattern~\cite{ofek,Zhang+09,halperin-rosenow} and the nature of the decoherence~\cite{heiblum+06,roulleau,McClure+09}. 

Quantum Hall interferometers are realized~\cite{heiblum} by current-carrying edge states~\cite{buttiker} (ES), arising at the intersections of the Fermi level and filled Landau levels at the sample edge. Usually, a key part of the interferometer scheme is a quantum point contact (QPC), which enables a connection between two identical counter-propagating edge states (ES)~\cite{stern}. By using two QPC in a proper sequence, an electronic analog of Mach-Zehnder~\cite{heiblum,heiblum+06,heiblum+07,litvin,litvin1,glattli,roulleau,roulleau1} or Fabry-Perot~\cite{wiel,ofek,Camino+05,Camino+07a,Camino+07b,Ping+09,Zhang+09,McClure+09} interferometer can be realized.

The interference pattern reflects the operation regime of quantum interferometer~\cite{halperin-rosenow}. In the simplest case of the extreme Aharonov-Bohm (AB) regime, electron-electron interaction has no effect on the interference pattern. The interference period corresponds to the change of the flux $\Phi=BS$ through the interferometer loop area $S$ by one flux quantum $\Phi_0$, where $S$ is practically independent of the magnetic field $B$.  In the opposite extreme Coulomb-dominated (CD) regime, $\Phi$ depends also on the number of particles within the loop because of Coulomb interaction~\cite{halperin-rosenow}. The interference pattern reveals complicated structures in $(B,S)$ plane in between these two extreme regimes~\cite{halperin-rosenow}. It was recently demonstrated~\cite{ofek,Zhang+09}, that the extreme CD and mixed AB-CD regimes are realized in small (about 2~$\mu$m size) devices~\cite{ofek}, while the extreme AB regime is realized for big (20~$\mu$m) QPC-based interferometers~\cite{Zhang+09}.

Different mechanisms were proposed to explain coherence~\cite{feldman,sukhorukov,chalker,neder,youn,levkivskiy} in the QPC-based interferometers. The coherence length was found~\cite{roulleau,roulleau1,McClure+09} to be inversely proportional to the temperature, which is compatible with theoretical predictions based on a dephasing arising from the thermal noise of the environment~\cite{roulleau1}. However, the complete theory is still missing for the dependence of the dephasing on   magnetic field and   imbalance~\cite{McClure+09}.

Visibility of the interference oscillations can be seriously suppressed even by low (of the order of microvolts) voltage imbalances in  QPC~\cite{wiel,heiblum+06,heiblum+07,litvin,litvin1,glattli,roulleau,roulleau1,McClure+09}. By contrast, a clear visible interference was reported even at millivolt imbalances for the interferometers realized by two  co-propagating ES at a single sample edge~\cite{fabry,fabryfrac}.  The evident discrepancy of the results implies a substantially different operation principle for the interferometers realized by co-propagating ES. 

Here, we experimentally realize quantum Hall Mach-Zehnder interferometer which operates far beyond the equilibrium. The operation of the interferometer is based on allowed intra-edge elastic transitions within the same Landau sublevel in the regime of high imbalances between the co-propagating edge states. Since the every edge state is definitely connected with the certain Landau sublevel, the formation of the interference loop can be understood as a splitting and a further reconnection of a single edge state. We observe an Aharonov-Bohm type interference pattern even for low-size interferometers. This novel interference scheme demonstrates high visibility even at millivolt imbalances and survives in a wide temperature range.

\section{Samples and technique}

\begin{figure}
\includegraphics*[width=0.75\columnwidth]{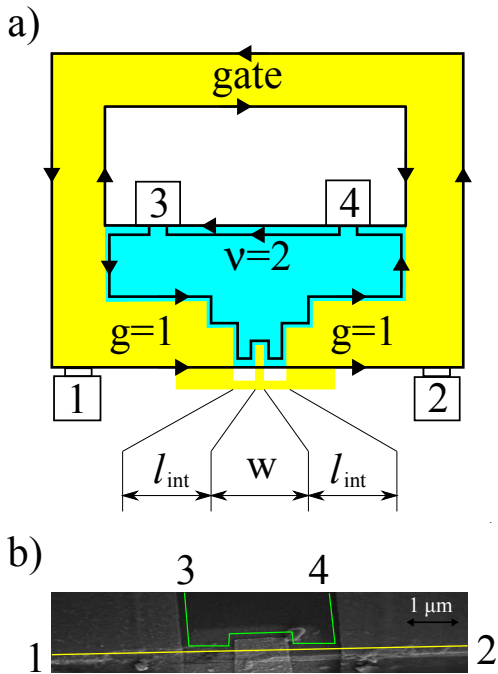}%
\caption{(Color online) (a) Schematic diagram of the sample (not to the scale). The outer sample dimension is about 2x2~mm$^2$. Each sample has a macroscopic ($\sim 0.5\times 0.5\mbox{mm}^2$) etched region inside. Ohmic contacts are placed at the mesa edges, denoted by bars with numbers. A split-gate (light yellow) partially encircles the etched area, leaving uncovered narrow gate-gap region at the outer mesa edge. Light green  area indicates uncovered 2DEG. The $w=$1 or 1.5~$\mu$m width gate finger is placed at the center of the gate-gap, being connected to the main gate outside the mesa. The lithographic overlap with the mesa is equal to $h=0.3 \mu$m. The gate finger separates two ES (thick lines) in the gate-gap, so the inter-ES transitions are only allowed in two interaction regions of widths $l_{int}=1 \mu$m. Arrows indicate an electron propagation direction within ES. (b) SEM image of the gate finger region. Two ES (green and yellow thick lines) are sketched as a straight lines regardless roughness of the mesa edge. For measurements one of the inner contacts (3 or 4) is grounded. Current is applied to one of the outer contacts (1 or 2), while the other outer contact is used to trace the outer ES potential. 
\label{sample}}
\end{figure}

Our samples are fabricated from a molecular beam epitaxially-grown GaAs/AlGaAs heterostructure. It contains a two-dimensional electron gas (2DEG) located 200~nm below the surface. The 2DEG mobility at 4K is  $5.5 \cdot 10^{6}  $cm$^{2}$/Vs  and the carrier density is   $1.43 \cdot 10^{11}  $cm$^{-2}$.

A novel sample design realizes a quantum Hall interferometer based on independently contacted co-propagating edge states, see Fig.~\ref{sample}. Edge states arise at the sample edges as the intersections of the Fermi level and filled Landau sublevels~\cite{buttiker}. At the bulk filling factor $\nu=2$, there are two co-propagating spin-split ES along the ungated mesa edges. The main gate redirects the inner ES, see Fig~\ref{sample} (a), by depleting 2DEG underneath to a lower filling factor $g=1$. It allows an independent contacting of two co-propagating ES in the gate-gap region at the outer mesa edge. The gate finger at the center of the  gate-gap region divides the inter-ES junction onto two ones. We study samples with two different gate finger widths $w=$1~$\mu$m or 1.5~$\mu$m.

To study the transmittance of the device we ground one of the inner Ohmic contacts (3 or 4) and apply a \textit{dc} current to one of the outer contacts (1 or 2). The other outer contact is used to trace the outer ES potential. We check that the obtained interference pattern does not depend on the particular choice of the  contact combination, which only affects the total resistance of the device. 

At positive currents, an electron contribute to transport only if it is transferred from the inner ES to the outer one in the gate-gap region. It can occur with some probability in the first ES junction of width $l_{int}=1 \mu$m or an electron can follow the inner ES around the gate finger and be transferred in the second one, see Fig.~\ref{sample}. In a very naive picture these two transmission regions serves as two semi-transparent mirrors in optical Mach-Zehnder interferometer, while two paths around the gate finger define the interferometer arms. This naive picture does not concern sophisticated dependence of the inter-ES transport on ES imbalance at a single sample edge, see for a review Ref.~\onlinecite{ESreview}. We discuss the interferometer operation in detail after the presenting of the experimental results. 

The phase difference $\phi=\Phi/\Phi_0$ between the interferometer arms is controlled by the flux $\Phi$ through the gate finger area. $\Phi$ can be affected~\cite{halperin-rosenow} by varying either the magnetic field $B$ or the effective gate finger area $S$. Within the QH plateau, $S$ is  sensitive to the {\em top} gate voltage   $V_g$ only because of varying the depletion width along the gate perimeter.  By contrast  to commonly used plunger gates, $S$ is increasing with  further depleting the top gate.

The measurements are performed in a dilution refrigerator with the minimal temperature of 30~mK. The interference pattern is independent of the cooling cycle. Standard two-point magnetoresistance is used to determine the regions of $B$ which correspond to integer QH filling factors $\nu$ in the ungated area. We measure capacitance between the gate and 2DEG as a function of the gate voltage $V_g$ at constant magnetic field to find $V_g$ regions of integer filling factors $g$ under the gate.

\section{Experimental results}

\begin{figure}
\includegraphics[width=\columnwidth]{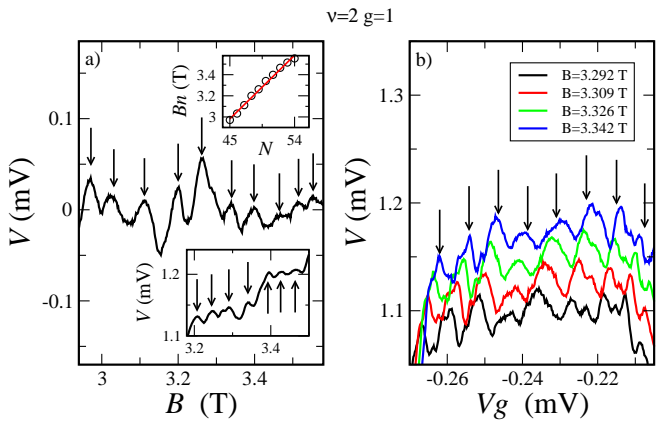}%
\caption{ (Color online) Examples of the oscillating behavior while sweeping the magnetic field  $B$  at constant gate voltage $V_g=-0.215$~V (a) or the gate voltage $V_g$ at several constant magnetic fields (b) within the $g=1$ QH state under the gate finger. Arrows indicate the positions of the oscillations. They are equidistant, see upper inset to part (a), and their evolution demonstrates the extreme AB interference regime (b). The obtained periods are $\Delta B=67$~mT and $\Delta V_{g}=7.8$~mV respectively.  A monotonous increase in $V(B)$ is subtracted from the $V(B)$ curve in the part (a). The curves in the part (b) remain unchanged, the signal drops down at the $g=1$ QH plateau edges (see also Fig.~\protect\ref{temp} (a)).     Measurement current is $I=4$~nA. Lower inset to part (a) demonstrates raw $V(B)$ oscillations with $\Delta B=45$~mT for $w=1.5 \mu$m sample, $I=10$~nA. Filling factors are $\nu=2, g=1$.    \label{oscill21}}
\end{figure}

\begin{figure}
\includegraphics[width=\columnwidth]{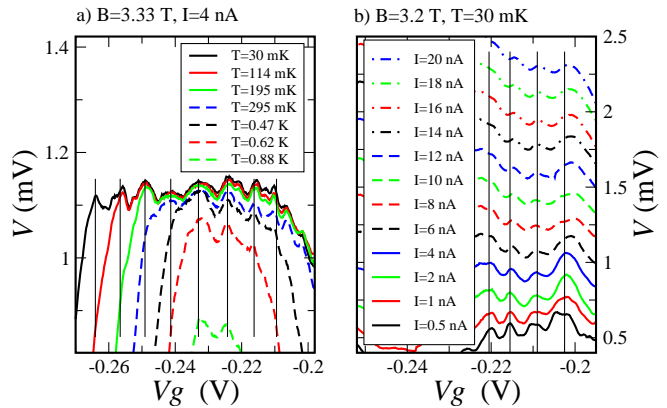}%
\caption{(Color online) (a) Oscillations in $V$ for different temperatures $T$. The oscillations amplitude is  insensitive to the temperature, in contrast to the width of the $g=1$ QH plateau and the mean level  of the signal.  The signal drops down at the $g=1$ QH plateau edges. Measurement current is $I=4$~nA, magnetic field is $B=3.33$~T. (b) Oscillations for different imbalances (measurement currents) at $B=3.2$~T, $T=30$~mK. The amplitude is practically independent of the imbalance. The mean value of the signal reflects the $I-V$ dependence (the lowest curves are still shifted vertically in the figure to avoid overlap). Thick lines indicate the positions of the oscillations, which differ in (a) and (b) because of different fields.   \label{temp}}
\end{figure}

To observe the interference effects in the transmittance of the device, we apply fixed dc current $I$ in the range of $0.5-20$~nA across the gate-gap junction as described before.  The potential of the outer contact $V(B,V_g)$ is shown in Fig.~\ref{oscill21}  in dependence on the magnetic field $B$ (a) or the gate voltage $V_g$ (b). Both $B,V_g$ are varied within the $g=1$ QH state under the gate finger. A monotonous increase in $V(B)$ is subtracted from the curve in Fig.~\ref{oscill21} (a). The curves in Fig.~\ref{oscill21} (b) remain unchanged. The $V(V_g)$ signal drops down at the edges of the $g=1$ QH plateau (see also Fig.~\protect\ref{temp} (a)), because at finite $\sigma_{xx}$ under the gate some part of the current flows outside the gate-gap region.

Both $V(B)$ and $V(V_g)$ dependencies  exhibit nearly equidistant (see inset to Fig.~\ref{oscill21} (a)) oscillations with periods $\Delta B=67$~mT and $\Delta V_{g}=7.8$~mV respectively. Fig.~\ref{oscill21} (b) also shows the evolution of the oscillation picture with increasing the magnetic field. The position of every oscillation moves to higher $V_g$ with increasing the magnetic field $B$. Since  the increase in $V_g$ lowers the depletion, it lowers $S$, so this behavior is a fingerprint of the extreme AB interference regime~\cite{halperin-rosenow} $\phi=$BS$/\Phi_0$.  This behavior is demonstrated for a wide magnetic field range.

Because of a simple relation $\phi=BS/\Phi_0$ in the extreme AB interferometer regime, $\Delta B$ reflects the whole active interferometer area $S$, $\Delta B=\Phi_0/S$. High values of the oscillation numbers $N=B/\Delta B$ in the inset to Fig.~\ref{oscill21} (a) also support the invariance of the loop area $S$ while changing the flux $\Phi$ on a quantum $\Phi_0$.  The  estimation $S=\Phi_0/\Delta B\approx 0.1\mu\mbox{m}^2$ is in a reasonable agreement with the gate finger dimensions, because the lithographic   length $h=0.3 \mu$m should be corrected by the a depletion length of roughly the 2DEG depth ($\approx 200$~nm) at the mesa edge~\cite{Zhang+09,McClure+09,shklovskii}. The sample with the bigger $w=1.5 \mu$m and demonstrates oscillations with $\Delta B=45$~mT  which corresponds to $S=\Phi_0/\Delta B\approx 0.15\mu\mbox{m}^2$. Thus, the period $\Delta B$ reasonably scales with the gate finger width $w$ even in view of the obvious roughness of the etched mesa edge. 

The above described picture of the extreme AB interference regime is also confirmed by measurements at other integer filling factors. Usually $\Delta B$ scales with filling factors for the small QPC-based interferometers~\cite{ofek}, which is a characteristic feature of the CD interference regime~\cite{halperin-rosenow}. In the present investigation, we obtain $\Delta B=70$~mT for $\nu=3, g=1$ and slightly higher $\Delta B=90$~mT for $\nu=3, g=2$ fillings. There are three ES in the gate-gap junction at the bulk filling factor $\nu=3$. Because at $g=1$ there is one ES under the gate finger, the interferometer geometry is formed by the same two outer spin-split ES as in the $\nu=2, g=1$ case. For $g=2$, $S$ is somewhat diminished because of two ES under the gate finger, which is reflected in a higher period.

By contrast to previous investigations~\cite{heiblum,heiblum+06,heiblum+07,litvin,litvin1,glattli,roulleau,ofek,Camino+05,Camino+07a,Camino+07b,Ping+09,Zhang+09,McClure+09}, the interference is certainly demonstrated at millivolt imbalances in Figs.~\ref{oscill21}. Moreover, the interference oscillations appears only above some threshold imbalance value. Above this value, they are nearly independent of the applied imbalance, see Fig.~\ref{temp} (b). The oscillations also remain practically unchanged with increasing the temperature in Fig.~\ref{temp} (a), while the $g=1$ QH plateau width, as determined by the signal drop, is highly sensitive to the temperature. The oscillations are visible even at $T=0.88$~K, where the plateau is very narrow and the level of the signal is diminished because of an admixture of the dissipative current under the gate finger, see Fig.~\ref{temp} (a).

\section{Discussion}

As a result, (i) we  undoubtedly observe the interference oscillations at millivolt imbalances in our device; (ii) the  visibility of the oscillations does not depend on the imbalance and temperature; (iii) the interferometer operates in the extreme AB regime, i.e. electron-electron interaction has minor effect on the interference pattern. This behavior is strongly different from the reported one for the interferometers based on QPC~\cite{heiblum,heiblum+06,heiblum+07,litvin,litvin1,glattli,roulleau,ofek,Camino+05,Camino+07a,Camino+07b,Ping+09,Zhang+09,McClure+09}. 

\begin{figure}
\centerline{\includegraphics[width=\columnwidth]{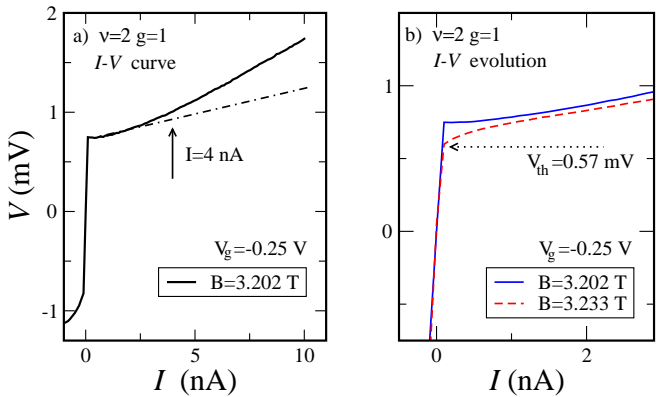}}%
\caption{  $I-V$ curves of the gate-gap junction for transport between two spin-split ES for the filling factors $\nu=2, g=1$. (a) The $I-V$ curve at constant $B=3.203$~T and $V_g=-0.25$~V in a wide current range. The positive branch decreases its slope abruptly at $V_{th}$, because of flattening the potential jump between two ES, cp. the energy diagrams in Fig.~\protect\ref{mz_diagram} (a) and (b). Dash-dot indicate the equilibrium  resistance $R_{eq}=2h/e^2$ above $V_{th}$. Highly non-linear negative branch is shown around the zero region only.  (b) Evolution of the experimental $I-V$ curve near $V_{th}$ with small variation of the magnetic field $B$. The value $V_{th}=0.57$~mV obtained at $B=3.233$~T coincides well with one for samples without gate finger, see Ref.~\onlinecite{zeeman}.  \label{IV}}
\end{figure}

\subsection{Evolution of $I-V$ characteristics}

To understand an origing of the interference oscillations at high imbalances, we consider $I-V$ characteristics of the gate-gap junction. 

Let us start from the $I-V$ curve at constant $B,V_{g}$. A typical experimental $I-V$ curve  is shown in  Fig.~\ref{IV} (a) for the filling factors $\nu=2, g=1$. At low imbalances, the positive $I-V$ branch is characterized  by high resistance, so the inter-ES transport is negligible. Above some threshold imbalance value $V_{th}$ the resistance drops significantly, which indicates much higher transport. The negative $I-V$ branch is  strongly non-linear and does not contain any specific points~\cite{ESreview}. 

Since the phase $\phi=\Phi/\Phi_0$ is practically independent of the imbalance at constant $B,V_{g}$, see Fig.~\ref{temp} (b), we can expect $I-V$ to be determined by the edge energy structure in the gate-gap junction, in complete similarity to the structures without the gate finger, see Refs.~\onlinecite{ESreview,zeeman}. The edge energy structure is depicted in  Fig.~\ref{mz_diagram}, (a), in the equilibrium~\cite{shklovskii}. In samples with a smooth edge profile, edge states~\cite{buttiker} are represented by  compressible strips of finite width~\cite{shklovskii}, located at the intersections of the Fermi level and filled Landau levels. 

Applied  inter-ES electrochemical potential imbalance modifies the potential jump between two ES. It is slightly diminished for low positive imbalances, because of the polarity in our setup. The inter-ES transport is still  negligible because the charge equilibration length~\cite{mueller} $l_{eq}\sim 100 \mu$m  exceeds significantly  the total  gate-gap junction width 2$l_{int}=2 \mu$m.  On the other hand, high positive inter-ES imbalance $V$ seriously distorts the edge energy structure~\cite{ESreview,zeeman}. It flattens the potential jump between two ES at some value $V=V_{th}$, see Fig.~\ref{mz_diagram} (b), which is reflected by the resistance drop in the experimental $I-V$ curve. At higher imbalances the potential between ES is even more distorted, so the resistance still exceeds it equilibrium value~\cite{ESreview,zeeman}, see Fig.~\ref{IV} (a).  

We want to emphasize that despite the theory~\cite{shklovskii} was developed only for the equilibrium, the diagrams in Fig.~\ref{mz_diagram} are qualitatively proven by the $I-V$ curves spectroscopy~\cite{ESreview,zeeman}. For the samples without the gate finger, $V_{th}$ simply reflects the Zeeman splitting at the edge, which is monotonically increasing with the magnetic field~\cite{zeeman}. This is the reason for the monotonous increase in $V(B)$ which is subtracted from the curve in Fig.~\ref{oscill21} (a). The value $V_{th}=0.57$~mV obtained from the $I-V$ curve at $B=3.233$~T coincides well with one for the samples without gate finger, see Ref.~\onlinecite{zeeman}.

An example of the $I-V$ evolution is shown in Fig.~\ref{IV} (b). We change the phase $\phi=\Phi/\Phi_0$ by  low variation of the magnetic field $B$ at the constant gate voltage. Two values of the magnetic field in Fig.~\ref{IV} (b) are close to the two neighbor maximum and minimum in Fig.~\ref{oscill21} (a).  The effect appears as a vertical shift of the curve at $V>V_{th}$.  For these two curves the shift is {\em negative} for the positive increase of the magnetic field, which is just opposite to the monotonous Zeeman increase of $V_{th}$. 

Fig.~\ref{IV} (b) is another representation of the interference independence of the applied imbalance in Fig.~\ref{temp} (b). It clearly demonstrates that   no interference effects can be seen below $V_{th}$ in our experimental setup.

\subsection{Interferometer operation}

\begin{figure}
\includegraphics*[width=\columnwidth]{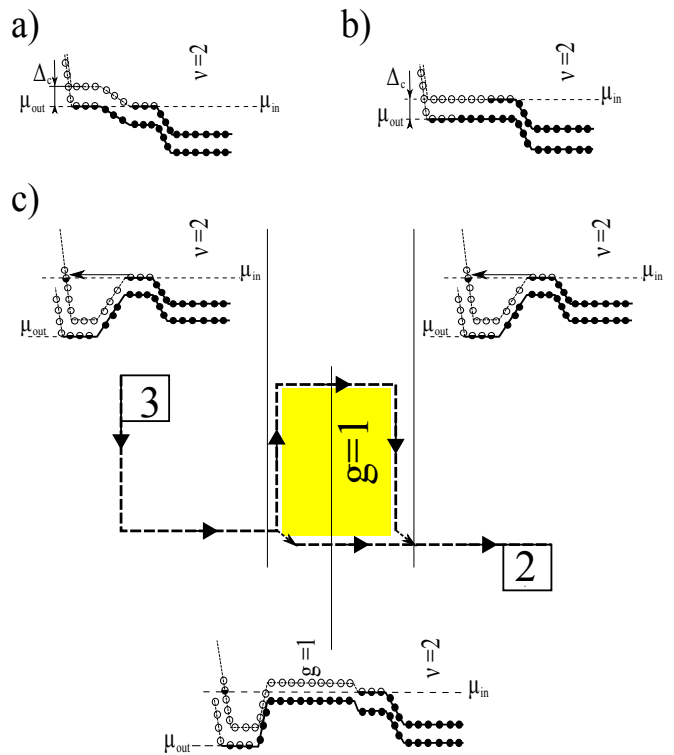}%
\caption{(Color online) (a) Energy diagram in the gate-gap region in the equilibrium~\protect\cite{ESreview} ($\mu_{out}=\mu_{in}$, no electrochemical potential imbalance is applied). Pinning~\protect\cite{shklovskii} of the energy levels to the Fermi level (shot-dash) is shown in the compressible regions (ES). Half-filled circles indicate partially occupied electron states in the compressible strips. Filled (open) circles represent the fully occupied (empty) electron states. $\Delta_c$ is the potential jump between two ES.   (b) Flattening of the potential jump by the positive voltage $V$, $eV=eV_{th}=\mu_{out}-\mu_{in}=-\Delta_c,e<0$. (c) An electron propagation around the interference loop in the gate finger area (light yellow).  Energy diagrams are shown before the gate finger (left), across it (bottom), and after the gate finger (right) at high imbalances $V>V_{th}$. Arrows indicate the intra-edge elastic transitions without spin-flip. 
\label{mz_diagram}}
\end{figure}

Let us concern the formation of the interference loop for an electron in detail. Fig.~\ref{mz_diagram}, (c), illustrates the energy diagrams for different positions within the gate-gap junction  in Fig.~\ref{sample}. 

If we consider the inter-ES transport to both sides of the gate finger, two major possibilities are allowed for an electron: (i) it can be directly transferred between the ground states in two ES, which is accompanied by the energy loss and a spin flip, i.e. by the coherence loss; (ii) at $V>V_{th}$ an electron can be elastically transferred within the same energy sublevel, see Fig.~\ref{mz_diagram} (c), with afterward relaxation to the ground state in the outer ES. The latter possibility preserves the coherence for the time scale smaller than the relaxation time. 

Thus, if the gate finger width $w$ is smaller than the relaxation length, an electron can cross the gate finger region by two paths, see Fig.~\ref{mz_diagram} (c): it encircles the finger gate along the inner ES or propagates along the outer mesa edge being in the excited state here (bottom diagram). These two paths form two  arms of the interferometer loop, being reconnected at the opposite finger edge. 

We worth to mention, that an electron placed at any interferometer arm still belongs to the same Landau sublevel. Since the every ES is definitely connected with the certain Landau sublevel, the formation of the interference loop can be understood as a splitting and a further reconnection of the inner ES. 

The proposed interferometer operation scheme is supported by the experimental facts. We observe the interference oscillations only in the regime of allowed elastic transitions $V>V_{th}$, see, e.g., Fig.~\ref{IV} (b). In this regime, an electron is at the same energy for both interferometer arms. This energy is independent of the exact value of the inter-ES imbalance, which only slightly affects the position of the outer arm. The drift velocity for every arm is determined by the energy level slope, which is also insensitive to the bias applied {\em between} two compressible strips, see Fig.~\ref{mz_diagram} (c). Thus, the inter-ES imbalance can only have a minor effect on the oscillations, as we observe in the experiment, see Fig.~\ref{temp} (b). The proposed scheme is also independent of the temperature, as long as the temperature is much below $eV_{th}$, as demonstrated in Fig.~\ref{temp} (a).

\subsection{Coherence}

In the proposed interferometer scheme, based on elastic intra-edge transitions, the coherence is firstly  restricted  by the  relaxation of the non-equilibrium electron in Fig.~\ref{mz_diagram}, (c). 

The energy relaxation length was studied in Ref.~\onlinecite{sueur} for low (microvolt) imbalances. It was found to exceed 10~$\mu$m, which is well above the gate finger widths $w=1, 1.5 \mu$m in our devices. On the other hand, the relaxation mechanism at high imbalances may be different.  We can only estimate it to exceed 5~$\mu$m as an indirect result of the $I-V$ curves spectroscopy investigations for samples with different gate-gap widths~\cite{ESreview}.

For the QPC-based interferometers, the coherence length investigations~\cite{roulleau,roulleau1,McClure+09} establish a dephasing to arise from the thermal charge noise of the environment~\cite{roulleau1}. However, the complete theory is still missing even for low imbalances~\cite{McClure+09}. From the present experiment we can not specify the decoherence mechanism, except for two statements: (i) the coherence is insensitive to the temperature as long as the temperature is much below $eV_{th}$, see Fig.~\ref{temp} (a); (ii) the decoherence mechanism connected with neutral collective modes~\cite{levkivskiy} might be important because of allowed emission of these modes at high imbalances, see Ref.~\onlinecite{neutral}.

\section{Conclusion}

As a conclusion, we experimentally realize quantum Hall Mach-Zehnder interferometer which operates far beyond the equilibrium. The operation of the interferometer is based on allowed intra-edge elastic transitions within the same Landau sublevel in the regime of high imbalances between the co-propagating edge states. Since the every edge state is definitely connected with the certain Landau sublevel, the formation of the interference loop can be understood as a splitting and a further reconnection of a single edge state. We observe an Aharonov-Bohm type interference pattern even for low-size interferometers. This novel interference scheme demonstrates high visibility even at millivolt imbalances and survives in a wide temperature range. 

\acknowledgments

We wish to thank  V.T.~Dolgopolov and D.E. Feldman for fruitful discussions.
We gratefully acknowledge financial support by the RFBR, RAS, the Program ``The State Support of Leading Scientific Schools''.


\begin{thebibliography}{99}




\bibitem{heiblum} Y.~Ji, Y.~Chung, D.~Sprinzak, M.~Heiblum, D.~Mahalu, and H.~Shtrikman, Nature {\bf 422}, 415 (2003).
\bibitem{heiblum+06} I. Neder, M. Heiblum, Y. Levinson, D. Mahalu, and V. Umansky, Phys. Rev. Lett. \textbf{96}, 016804 (2006).
\bibitem{heiblum+07} I. Neder, M. Heiblum, D. Mahalu, and V. Umansky, Phys. Rev. Lett. \textbf{98}, 036803
 (2007).
\bibitem{litvin}  L. V. Litvin, H.-P. Tranitz, W. Wegscheider, and C. Strunk, Phys. Rev. B \textbf{75}, 033315 (2007).
\bibitem{litvin1}  L. V. Litvin, A. Helzel, H.-P. Tranitz, W. Wegscheider, C. Strunk, arXiv:0802.1164
\bibitem{glattli} Preden Roulleau, F. Portier, D. C. Glattli, P. Roche, A. Cavanna, G. Faini, U. Gennser, and D. Mailly  Phys. Rev. B \textbf{76}, 161309(R) (2007).
\bibitem{roulleau} P. Roulleau, F. Portier, P. Roche, A. Cavanna, G. Faini, U. Gennser, D. Mailly, Phys. Rev. Lett. \textbf{100}, 126802 (2008)
\bibitem{roulleau1} P. Roulleau, F. Portier, P. Roche, A. Cavanna, G. Faini, U. Gennser, and D. Mailly, Phys. Rev. Lett. 101, 186803 (2008).

\bibitem{wiel}  W. G. van der Wiel, Y. V. Nazarov, S. DeFranceschi, T. Fujisawa, J. M. Elzerman, E. W. G. M. Huizeling, S. Tarucha, and L. P. Kouwenhoven, Phys. Rev. B {\bf 67},  033307 (2003).
\bibitem{ofek} N.~Ofek, A.~Bid, M.~Heiblum, A.~Stern, V.~Umansky, D.~Mahalu, PNAS 107, 5276-5281 (2010).
\bibitem{Camino+05} F.E.~Camino, Wei Zhou, and V.J.~Goldman, Phys.
Rev. Lett. {\bf 95}, 246802 (2005).
\bibitem{Camino+07a} F.E.~Camino, Wei Zhou, and V.J.~Goldman, Phys. Rev. B {\bf 76}, 155305 (2007).
\bibitem{Camino+07b} F.E.~Camino, Wei Zhou, and V.J.~Goldman, Phys. Rev. Lett. {\bf 98}, 076805 (2007).
\bibitem{Ping+09} Ping V.~Lin, F.E.~Camino, and V.J.~Goldman, Phys. Rev. B {\bf 80}, 125310 (2009).
\bibitem{Zhang+09} Y.~Zhang, D.T.~McClure, E.M.~Levenson-Falk, C.M.~Marcus, L.N.~Pfeiffer, and K.W.~West,
Phys. Rev. B {\bf 79},  241304 (2009).
\bibitem{McClure+09} D.T.~McClure, Y.~Zhang, B.~Rosenow, E.M.~Levenson-Falk, C.M.~Marcus, L.N.~Pfeiffer, and K.W.~West, Phys. Rev. Lett. {\bf  103}, 206806 (2009).

\bibitem{halperin-rosenow}B.I.~Halperin, A.~Stern, I.~Neder, B.~Rosenow, arXiv:1010.4598

\bibitem{buttiker} M. B\"uttiker, Phys. Rev. B {\bf 38}, 9375 (1988).
\bibitem{stern} For a review see A.~Stern, arXiv:0711.4697

\bibitem{feldman} K. T. Law, D. E. Feldman, and Yuval Gefen, Phys. Rev. B 74, 045319 (2006)
\bibitem{sukhorukov} E. V. Sukhorukov and V. V. Cheianov, Phys. Rev. Lett. 99, 156801 ͑2007͒.
\bibitem{chalker} J. T. Chalker, Y. Gefen, and M. Y. Veillette, Phys. Rev. B 76,
085320 ͑2007͒.
\bibitem{neder} I. Neder and E. Ginossar, Phys. Rev. Lett. 100, 196806 ͑2008͒.
\bibitem{youn} S.-C. Youn, H.-W. Lee, and H.-S. Sim, Phys. Rev. Lett. 100,
196807 ͑2008͒.
\bibitem{levkivskiy} Ivan P. Levkivskyi and Eugene V. Sukhorukov, Phys. Rev. B 78, 045322 (2008)






\bibitem{fabry} E.V. Deviatov and A. Lorke, Phys. Rev. B \textbf{77}, 161302(R) (2008)
\bibitem{fabryfrac} E. V. Deviatov, B. Marquardt, A. Lorke, G. Biasiol, and L. Sorba, Phys. Rev. B \textbf{79}, 125312 (2009). 




\bibitem{shklovskii} D. B. Chklovskii, B. I. Shklovskii, and L. I. Glazman, Phys. Rev. B {\bf 46}, 4026 (1992).
\bibitem{mueller}G. M\"uller, D. Weiss, A. V. Khaetskii,  K. von Klitzing,  S. Koch, H. Nickel, W. Schlapp, and R. L\"osch,  Phys. Rev. B {\bf 45}, 3932 (1992).
\bibitem{ESreview} For a review on inter-ES transport see E. V. Deviatov, A Lorke, phys. stat. sol. (b) 245, 366 (2008).
\bibitem{zeeman} E.V. Deviatov, A. Lorke, G. Biasiol, L. Sorba, W. Wegscheider, JETP Lett. {\bf 92},  69 (2010).


\bibitem{sueur} H. le Sueur, C. Altimiras, U. Gennser, A. Cavanna, D. Mailly, F. Pierre, Phys. Rev. Lett. 105, 056803 (2010).
\bibitem{neutral} E. V. Deviatov, A. Lorke, G. Biasiol, and L. Sorba, Phys. Rev. Lett. 106, 256802 (2011).

\end{thebibliography}
\end{document}